\documentclass{siamltex}

\usepackage{amsmath, latexsym, amsfonts, amssymb, amscd,bbm}
\usepackage[english]{babel}
\usepackage[dvips]{graphicx}
\usepackage[mathlines]{lineno}

\usepackage{amsmath,amsfonts,bm,bbm}

\usepackage{float}
\usepackage{epsfig}
\usepackage{esint}
\usepackage{color}

\usepackage{multirow}

  \usepackage{paralist}
  \usepackage{epstopdf}
  \usepackage{graphics} 
 \usepackage[colorlinks=true]{hyperref}
 \hypersetup{urlcolor=blue, citecolor=red}
 \usepackage[latin1]{inputenc}
\usepackage[caption=false]{subfig}

\newcommand{\dis}{\displaystyle}

\usepackage{float}
\usepackage{epsfig}
\usepackage{esint}

\usepackage{bm,braket}

\newcommand{\calG}{{\mathcal G}}

 \newcommand{\calA}{{\mathcal A}}
 \newcommand{\calU}{{\mathcal U}}
\newcommand{\R}{{\mathbb R}}

\newcommand{\X}{\mathbf{X}}

\newcommand{\x}{\mathbf{x}}
\newcommand{\bx}{\overline{\mathbf{x}}}
\newcommand{\y}{\mathbf{y}}
\newcommand{\e}{{\mathrm e}}

\newcommand{\n}{\mathbf n}
\newcommand{\calT}{{\mathfrak T}}
\newcommand{\calQ}{{\mathcal Q}}
\newcommand{\calM}{{\mathfrak M}}
\newcommand{\calF}{{\mathcal F}}
\newcommand{\calJ}{{\mathcal J}}
\newcommand{\wphi}{\widetilde{\phi}}
\newcommand{\wrho}{\widetilde{\rho}}

\newcommand{\q}{\widetilde{q}}

\newcommand{\J}{\widetilde{J}}

\newcommand{\z}{\mathbf{z}}

\newcommand{\hxi}{{\bm \xi}}

\newcommand{\calP}{{\mathcal P}}

\newcommand{\p}{\widetilde{p}}

\title{Asymptotic analysis of the 2D narrow-capture problem for partially accessible targets}
\author{Paul C. Bressloff\thanks{Department of Mathematics, Imperial College London, London SW7 2AZ, UK  ({\tt p.bressloff@imperial.ac.uk}).}}

\begin{document}

\maketitle

\begin{abstract} 
In this paper we use singular perturbation theory to solve the 2D narrow capture problem for a set of partially accessible targets $\calU_k$, $k=1,\ldots,N$, in a bounded domain $\Omega\subset \R^2$. In contrast to previous models of narrow capture, we assume that when a searcher finds a target by attaching to the partially adsorbing surface $\partial \calU_k$ it does not have immediate access to the resources within the target interior. Instead, the searcher remains attached to the surface for a random waiting time $\tau$, after which it either gains access to the resources within ({\em surface absorption}) or detaches and continues its search process ({\em surface desorption)}. We also consider two distinct desorption scenarios -- either the particle continues its search from the point of desorption or rapidly returns to its initial search position. In applications to animal foraging, the latter would correspond to an animal returning to its home base whereas the resources within a target could represent food or shelter. We formulate the narrow capture problem in terms of a set of renewal equations that relate the probability density and target flux densities for absorption to the corresponding quantities for irreversible adsorption. The renewal equations, which effectively sew together successive rounds of adsorption and desorption prior to the final absorption event, provide a general probabilistic framework for incorporating non-Markovian models of desorption/absorption and different search scenarios following desorption. We solve the general renewal equations in two stages. First, we calculate the Laplace transformed target fluxes for irreversible adsorption by solving a Robbin boundary value problem (BVP) 
 in the small-target limit using matched asymptotic analysis. We then use the inner solution of the BVP to solve the corresponding Laplace transformed renewal equations for non-Markovian desorption/absorption, which leads to explicit Neumann series expansions of the corresponding 
target fluxes. Finally, the latter are used to determine the corresponding splitting probabilities and conditional mean FPTs for absorption.
\end{abstract}
\maketitle


\section{Introduction}

A classical stochastic search problem is an animal foraging for food or shelter located at some hidden target $\calU$ within a bounded domain $\Omega \subset \R^d$ \cite{Bell91,Bartumeus09,Viswanathan11}. The animal is said to find the target when it reaches the target boundary $\partial \calU$ for the first time. Similarly, in cell biology many biochemical reactions are triggered by a signaling molecule binding to the surface of some subscellular compartment $\calU$ within the cellular interior $\Omega$ \cite{Loverdo08,Benichou10,Bressloff13}. Mathematically speaking, target search is usually mapped to a first passage time (FPT) problem \cite{Greb24}. If the position of the searcher or particle at time $t$ is denoted by $\X(t)$, then the search process is terminated at the stopping time or FPT defined by $T=\inf\{t>0, \X(t) \in \partial \calU\}$. In many cases, there are multiple targets within
the interior of the search domain, which requires determining the splitting probability of being captured by a specific target. Since this probability is less than unity, it follows that the corresponding FPT density has infinite moments, unless it is conditioned on the set of events that find the target. The classical narrow capture problem involves a diffusive search process where the targets are much smaller than the size of the search domain. This then allows matched asymptotic expansions and Green's functions to be used to solve the corresponding boundary value problems (BVPs) for the splitting probabilities and moments of the conditional FPT density \cite{Bressloff08,Coombs09,Chevalier11,Coombs15,Ward15,Lindsay17}. These BVPs can be derived from the backward evolution equation for the probability density of particle position. Alternatively, one can determine statistical quantities of interest by solving the corresponding forward evolution equation in Laplace space \cite{Lindsay16,Bressloff21A,Bressloff21B}. This generates the Laplace transformed probability flux into each target surface, which acts as the FPT moment generator. Narrow capture is one example of a general class of singularly-perturbed diffusion problems with small interior traps, which have a wide range of applications in cell biology and beyond \cite{Ward18,Bressloff24}.

We have recently extended the classical single-target search problem for a Brownian particle by taking into account what may happen once the particle reaches the target surface \cite{Bressloff25,Bressloff25a}. First, rather than instantaneously binding or attaching to the surface, the particle may be reflected, resulting in an alternating sequence of bulk diffusion interspersed with local surface interactions prior to binding. That is, $\partial \calU$ acts as a {\em partially adsorbing} surface. Second, once the particle is adsorbed it may subsequently unbind and return to diffusion in the bulk ({\em desorption}) or be permanently transferred to the interior $\calU$ ({\em absorption}). We will refer to $\calU$ as a {\em partially accessible target}. Within the context of animal foraging, the inclusion of desorption and absorption means that the resources within $\calU$ are not immediately accessible and the animal has to find some way to penetrate the surface \cite{Bressloff25a}. Absorption and desorption are also a common feature of signal transduction in biological cells \cite{Bressloff21B}. That is, a signaling molecule (ligand) reversibly binds to the surface membrane. Moreover, in order to trigger a downstream signalling cascade, the bound molecule (and possibly its cognate receptor) has to be internalized via an active process known as endocytosis. A final element of the search process is specifying what happens after desorption. One possibility is that the particle continues diffusing from the point $\y \in \partial \calU_k$ at which it unbinds from the $k$th target. An alternative scenario is that the particle rapidly returns to its initial location $\x_0$. In the case of a foraging animal, this could represent its return to home base after failing to gather resources \cite{Bressloff25a}.

\begin{figure}[t!]
\begin{center} 
\includegraphics[width=12cm]{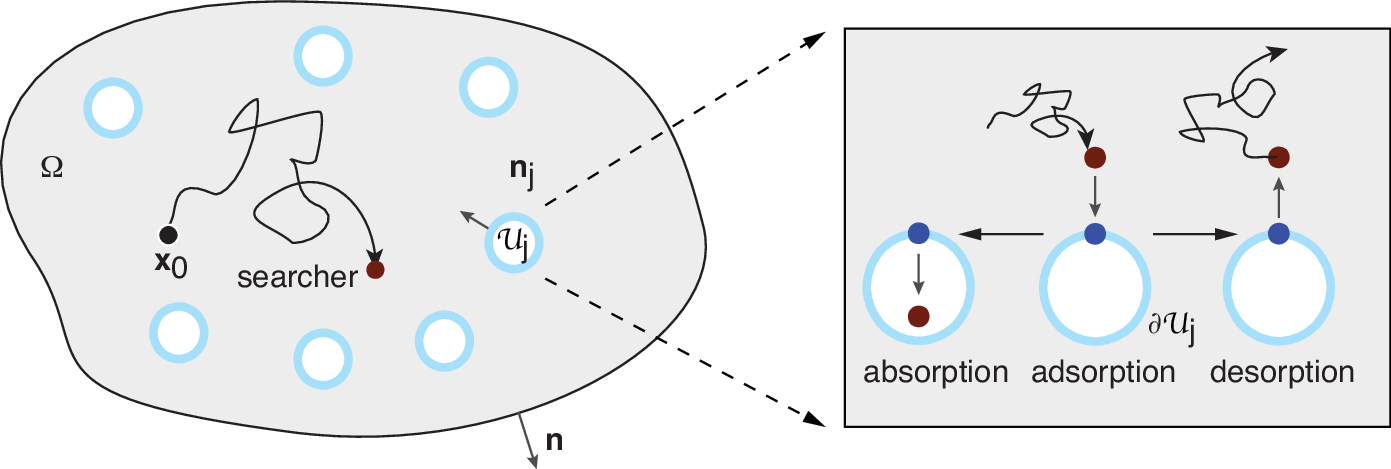} 
\caption{Diffusion of a searcher (particle)  in a bounded domain $\Omega$ with $N$ partially accessible targets $\calU_j$, $j=1,\ldots,N$. Each boundary surface $\partial \calU_j$ is taken to be partially adsorbing. However, adsorption of the particle at a point on $\partial \calU_j$ does not give it immediate access to the resources within $\calU_j$. After some random waiting time attached to the surface, the particle either succeeds in entering the interior $\calU_j$ (absorption) or detaches from the target to continue its search (desorption).}
\label{fig1}
\end{center}
\end{figure}

In this paper we analyze the 2D narrow capture problem in the case of $N$ partially accessible targets as illustrated in Fig. \ref{fig1}. Generalizing our recent work on the search for a single partially accessible target \cite{Bressloff25,Bressloff25a}, which itself builds upon a previous renewal formulation of reversible adsorption in physical chemistry \cite{Grebenkov23}, we construct a set of renewal equations that relate the probability density and target flux densities for absorption to the corresponding quantities for irreversible adsorption. The renewal equations effectively sew together successive rounds of adsorption and desorption prior to the final absorption event. One advantage of the renewal approach is that it is straightforward to incorporate non-Markovian models of absorption and desorption by taking the waiting time density for the duration of an adsorbed state prior to desorption/absorption to be non-exponential. The Markovian exponential case is equivalent to taking constant rates of desorption and absorption -- the probability density for particle position then satisfies a generalized Robin boundary-value problem (BVP), which couples to a set $N$ local variables that represent the probabilities that the particle is bound to one of  the target surfaces. A second useful feature of the renewal approach is that one can consider different desorption scenarios by modifying the rule for sewing together successive rounds of search-and-capture. In particular, rather than continuing the search process from the point of desorption, the particle may rapidly return to its initial position following desorption.

The structure of the paper is as follows. In \S 2, we formulate the generalized Robin BVP for Markovian desorption/absorption whose solution directly determines the target fluxes ${\calJ}_k(\x_0,t)$, $k=1,\ldots,N$. We then construct the renewal equations for the corresponding non-Markovian case. Each flux ${\calJ}_k(\x_0,t)$ now satisfies an integral renewal equation involving the corresponding set of fluxes $J_i(\x_0,t)$, $i=1,\ldots,N$, for irreversible adsorption -- the latter are defined in terms of the solution to a classical Robin BVP. If the particle returns to $\x_0$ after desorption, then the Laplace transformed renewal equations can be solved explicitly in the sense that $\widetilde{\calJ}_k(\x_0,s)$ is an explicit function of $\J_i(\x_0,s)$, $i=1,\ldots,N$. This no longer holds in the case of continuous search after desorption, since the corresponding renewal equations include non-trivial spatial integrals over the target surfaces.

 In \S 3 we derive explicit expressions for the Laplace transformed fluxes $\widetilde{\calJ}_k(\x_0,s)$ in the small-target limit using methods from singular perturbation theory. 
First, we use matched asymptotic analysis to solve the Laplace transformed Robin BVP for the fluxes $\J_i(\x_0,s)$, $i=1,\ldots,N$. Proceeding along similar lines to Ref. \cite{Bressloff21A}, we construct an inner or local solution
valid in an $O(\epsilon)$ neighborhood of each target, and then match to an outer or global solution that is valid away from each neighborhood. Here $\epsilon$ is a small dimensionless parameter that characterizes the size of each target relative to the size of the search domain. Since the 2D Neumann Green's function of the diffusion equation has a logarithmic singularity, the resulting asymptotic expansion is in powers of $\nu= -1/\ln \epsilon$ rather than $\epsilon$ itself. As originally shown in Refs. \cite{Ward93,Ward93a}, it is possible to sum over the logarithmic terms non-perturbatively by inverting a matrix with coefficients that are linear in $\nu$. This is equivalent to calculating the asymptotic solution for all terms of $O(\nu^k)$ for any $k$. Second, we extend the asymptotic analysis to the generalized Robin BVP for Markovian desorption/adsorption and derive an explicit Neumann series expansion of $\widetilde{\calJ}_k(\x_0,s)$ in terms of products of the fluxes $\J_i(\x_0,s)$, $i=1,\ldots,N$. The $n$th term in the expansion represents the contribution from $n-1$ rounds of desorption prior to absorption. Third, we use the inner solution of the Robin BVP to solve the Laplace transformed integral renewal equations for non-Markovian desorption/adsorption This yields generalized Neumann series expansions for the fluxes $\widetilde{\calJ}_k(\x_0,s)$, which reduce to the Markovian versions in the case of constant rates of desorption and absorption. Moreover,  the Neumann series can be formally summed to yield expressions for $\widetilde{\calJ}_k(\x_0,s)$ that are non-perturbative with respect to $\nu$.
Finally, in \S 4, we apply the results of our asymptotic analysis to determine the splitting probabilities $\pi_k(\x_0)$ and conditional MFPTs $\calT_k(\x_0)$ for absorption. We use the well-known identities $\pi_k(\x_0)=\lim_{s\rightarrow 0}\widetilde{\calJ}_k(\x_0,s)$ and $\pi_k(\x_0)\calT_k(\x_0)=-\lim_{s\rightarrow 0}\partial_s\widetilde{\calJ}_k(\x_0,s)$. We proceed by performing a small-$s$ expansion of the Neumann series derived in \S 3, which requires dealing with the singular nature of the Green's function in the limit $s\rightarrow 0$. We illustrate the theory by considering the simple example of a a pair or targets in the unit disc.

\vfill

 \section{Search for partially accessible targets}
 
 \subsection{Markovian desorption}
 
 Consider a Brownian particle that is diffusing in a bounded domain $\Omega\subset \R^2$ containing a set of $N$ partially accessible targets $\calU_k$, $k=1,\ldots,N$, with $\bigcup_{j=1}^N \calU_k=\calU_a\subset \Omega$, see Fig. \ref{fig1}. Whenever the particle hits a point $\y \in \partial \calU_j$, it either reflects or enters a bound state at a constant rate $\kappa_{0}$. (For simplicity the  adsorption rate is taken to be target-independent.) Suppose that the bound particle then either unbinds from the point $\y \in \partial \calU_j$ at a rate $ \gamma_j$ and continues diffusing or is permanently absorbed into the interior $\calU_j$ at a rate $\overline{\gamma}_j$. 
Let $\rho(\x,t|\x_0)$ be the probability density that at time $t$ a particle is at $\X(t)=\x$, having started at position $\x_0$. Similarly, let $q_j(\y,t|\x_0)$ denote the probability that the particle is bound to a point $\y\in \partial \calU_j$ at time $t$. Then
\begin{subequations} 
\label{master}
\begin{align}
	\frac{\partial \rho(\x,t|\x_0)}{\partial t} &= D{\bm \nabla}^2 \rho(\x,t|\x_0), \ \x\in \Omega\backslash \calU_a,\quad {\bm \nabla} \rho \cdot \n=0, \ \x\in \partial \calU,\\
	D{\bm \nabla} \rho(\y,t|\x_0)\cdot \n_j(\y)&=\kappa_0\rho(\y,t|\x_0) -\gamma_j q_j(\y,t|\x_0), \quad \y \in \partial \calU_j,
\end{align}
 and
\begin{equation}
\frac{\partial q_j(\y,t|\x_0) }{\partial t}=\kappa_0\rho(\y,t|\x_0) - (\gamma_j+\overline{\gamma}_j) q_j(\y,t|\x_0),\quad  \y \in \partial \calU_j,
\end{equation}
	\end{subequations} 
together with the initial condition $\rho(\x,t|\x_0)=\delta(\x-\x_0)$ and $q_j(\y,0|\x_0)=0$. Here $\n$ and $\n_j$ denote the outward normals on $\partial \Omega$ and $\partial \calU_j$, respectively.
 
 In the limit $\gamma_j\rightarrow 0$ (zero desorption), equation (\ref{master}b) reduces to the classical Robin boundary condition and adsorption is irreversible. On the other hand, if $\gamma_j>0$ then either adsorption is reversible ($\overline{\gamma}_j=0$) or partially reversible ($\overline{\gamma}_j>0$). If the latter holds, then the particle is ultimately absorbed into the interior of one of the targets. The probability that the particle is absorbed by the $k$-th target after time $t$ is
\begin{equation}
\label{Pi}
\Pi_k(\x_0,t)=\int_t^{\infty}\calJ_k(\x_0,\tau)d\tau,
\end{equation}
where
\begin{equation}
\calJ_k(\x_0,\tau)=\overline{\gamma}_k \int_{\partial \calU_k}q_k(\y,\tau|\x_0)d\y
\label{abJk}
\end{equation}
is the absorption flux across the surface $\partial \calU_k$.
The corresponding splitting probability is
\begin{equation}
\label{split}
\pi_k(\x_0)=\Pi_k(\x_0,0)=\int_0^{\infty}\calJ_k(\x_0,t)dt =\widetilde{\calJ}_k(\x_0,0),
\end{equation}
where $\widetilde{\calJ}_k(\x_0,s)=\int_0^{\infty}\e^{-st}\calJ_k(\x_0,t)dt$. In addition, the survival probability that the particle has not yet been absorbed by any target (irrespective of whether it is freely diffusing or bound to one of the target surfaces) is
\begin{equation}
\label{Qsur}
\calQ(\x_0,t)=1-\sum_{k=1}^N\int_0^t\calJ_k(\x_0,\tau)d\tau .
\end{equation}
Differentiating both sides with respect to time and using equations (\ref{master}b,c) yields
\begin{align}
\frac{d\calQ(\x_0,t)}{dt}&=-\sum_{k=1}^N\overline{\gamma}_k\int_{\partial \calU_k}q_k(\y,\tau|\x_0)d\y\nonumber \\
&=\sum_{k=1}^N\int_{\partial \calU_k}\bigg [\frac{\partial q_k(\y,t|\x_0)}{\partial t}-D{\bm \nabla} \rho(\y,t|\x_0)\cdot \n_j(\y)\bigg ]d\y.
\end{align}
Integrating equation (\ref{master}a) with respect to $\x\in \Omega\backslash \calU_a$ and using the divergence theorem shows that
\begin{align}
\frac{d\calQ(\x_0,t)}{dt}
&=\frac{d}{dt}\bigg [\sum_{k=1}^N\int_{\partial \calU_k}q_k(\y,t|\x_0)d\y+\int_{\Omega\backslash \calU_a} \rho(\x,t|\x_0)d\x \bigg ] .
\end{align}
If $\overline{\gamma}_k=0$ for all $k$ then $d\calQ(\x_0,t)/dt=0$ for $t\geq 0 $ and we have conservation of total probability:
\begin{equation}
\sum_{k=1}^N\int_{\partial \calU_k}q_k(\y,t|\x_0)d\y+\int_{\Omega\backslash \calU_a} \rho(\x,t|\x_0)d\x =1.
\end{equation}
On the other hand, if $\overline{\gamma}_j>0$ for at least one target $j$, then $\calQ(\x_0,t)\rightarrow 0$ as $t\rightarrow \infty$ (assuming $\Omega$ is bounded) and
\begin{equation}
\sum_{j=1}^N\pi_j(\x_0)=1.
\end{equation}

Finally, the unconditional FPT density for absorption by any of the targets is
\begin{equation}
\calF(\x_0,t)=-\frac{d\calQ(\x_0,t)}{dt}=\sum_{k=1}^N \calJ_k(\x_0,\tau).
\end{equation}
We can also set $\calJ_k(\x_0,\tau)=\pi_k\calF_k(\x_0,\tau)$ with $\calF_k(\x_0,\tau)$ the conditional FPT density for the $k$th target.
The Laplace transform of the FPT density, $\widetilde{\calF}(x_0,s)$, is the moment generating function with
 \begin{align}
 \calT^{(n)}(x_0)&:=\int_0^{\infty}t^n \calF(x_0,t)dt 
 =\left . \left (-\frac{d}{ds}\right )^n \widetilde{\calF}(x_0,s)\right |_{s=0},
 \end{align}
We conclude that the splitting probabilities and FPT densities can be obtained by solving the BVP (\ref{master}) in Laplace space
 The Laplace transformed probability density $\wrho$ is the solution of the Robin-like BVP
 \begin{subequations}
 \label{LTRobin}
\begin{align}
&D{\bm \nabla}^2 \wrho(\x,s|\x_0)-s\wrho(\x,s|\x_0)=-\delta(\x-\x_0),\quad \x \in \Omega \backslash \calU_a, \\
 &D{\bm \nabla}\wrho(\y,s|\x_0)\cdot \n_j(\y)=\kappa_0g_j(s)\wrho(\y,s|\x_0),\ \y\in \partial \calU_j, \\
 &\q_j(\y,s|\x_0)=\frac{\kappa_0}{s+\gamma_j+\overline{\gamma}_j}\wrho(\y,s|\x_0),\ \y\in \partial \calU_j,
\end{align}
\end{subequations}
with
\begin{equation}
\label{gs}
g_j(s)=\frac{(s+\overline{\gamma}_j)}{s+\gamma_j+\overline{\gamma}_j}.
\end{equation}
Moreover,
\begin{equation}
\widetilde{\calJ}_j(\x_0,\tau)=\frac{\kappa_0\overline{\gamma}_j}{s+\gamma_j+\overline{\gamma}_j} \int_{\partial \calU_j}\wrho(\y,s|\x_0)d\y.
\label{MaJ}
\end{equation}

\subsection{Non-Markovian desorption}

 Following Refs. \cite{Grebenkov23,Bressloff25,Bressloff25a}, we now consider a more general non-Markovian model of desorption namely, when the particle is adsorbed by the $j$th surface $\partial \calU_j$, it remains bound for a random time $\tau$ generated from a waiting time density $\phi_j(\tau)$. The particle then either desorbs with a splitting probability $\sigma_{j}$ or is permanently absorbed with probability $1-\sigma_{j}$. In the exponential case 
 \begin{equation}
 \label{ephi}
 \phi_j(\tau)=(\gamma_j+\overline{\gamma}_j)\e^{-(\gamma_j+\overline{\gamma}_j) \tau},
 \end{equation}
 with the associated desorption probability $\sigma_j=\gamma_j/(\gamma_j+\overline{\gamma}_j)$, we recover the Markovian kinetic scheme of constant rates of absorption and desorption as in equations (\ref{master}b,c). A general probabilistic framework for analyzing the diffusive search for a single partially accessible target has recently been developed using renewal theory \cite{Bressloff25,Bressloff25a}. (The corresponding renewal equations for reversible desorption were developed in Ref. \cite{Grebenkov23}.) The renewal equations relate the densities $\rho(\x,t|\x_0)$ and $\calF(\x_0,t)$ in the presence of desorption and absorption (partially reversible adsorption) to the corresponding quantities without desorption (irreversible adsorption).
We begin by writing down the analogous renewal equations for multiple targets:
\begin{subequations}
\label{2Dren}
\begin{align}
 \rho(\x,t|\x_0)&=p(\x,t|\x_0)\\
 &+\int_0^td\tau' \int_{\tau'}^t d\tau\, \sum_{k=1}^N \sigma_{k}\phi_k(\tau-\tau') \bigg [\int_{\partial \calU_k}  \rho(\x,t-\tau|\y)J_k(\y,\tau'|\x_0)d\y \bigg ],\nonumber \\ 
 \calJ_j(\x_0,t)=& \int_0^td\tau(1-{\sigma}_j){\phi}_j(t-\tau)  \bigg [\int_{\partial \calU_j} d\y\,   J_j(\y,\tau'|\x_0)\bigg ]. \\
  &\quad  +  \int_0^td\tau' \int_{\tau'}^t d\tau\, \sum_{k=1}^N \sigma_k\phi_k(\tau-\tau') \bigg [\int_{\partial \calU_k}\calJ_j(\y,t-\tau)J_k(\y,\tau'|\x_0)d\y\bigg ].\nonumber
 \end{align}
 \end{subequations}
 Here $p(\x,t|\x_0)$ is the probability density for irreversible adsorption and $J_j(\y,t|\x_0)$ is the corresponding adsorption flux density at a point $\y \in \partial \calU_j$. The density $p(\x,t|\x_0)$ satisfies the Robin BVP
 \begin{subequations}
 \label{master0}
 \begin{align}
	&\frac{\partial p(\x,t|\x_0)}{\partial t} = D\nabla^2 p(\x,t|\x_0), \ \x\in \Omega\backslash \calU_a,\quad {\bm \nabla} p \cdot \n=0, \ \x\in \partial \Omega,\\
	&J_j(\y,t|\x_0)\equiv D{\bm \nabla}p(\y,t|\x_0)\cdot \n_j(\y)=\kappa_0p(\y,t|\x_0) , \quad \y \in \partial \calU_j.
\end{align}
\end{subequations}
Also note that the total adsorption flux into the $j$th target satisfies
 \begin{equation}
 J_j(\x_0,t)=\int_{\partial \calU_j}J_j(\y,t|\x_0)d\y
 \end{equation}
and $f(\x_0,t)=\sum_{j=1}^NJ_j(\x_0,t)$ is the unconditional FPT density for adsorption by any one of the targets. 
The first term on the right-hand side of the renewal equation (\ref{2Dren}a) represents the contribution from all sample paths that start at $\x_0$ and have not been adsorbed over the interval $[0,t]$. The $k$th contribution on the second line represents all sample paths starting from $\x_0$ that are first adsorbed at some point $\y \in \partial \calU_k$  with flux density $J_k(\y,\tau'|\x_0)$, remain in the bound state until desorbing in the time interval $[\tau,\tau+d\tau]$ with probability $\sigma_k \phi_k(\tau-\tau')d\tau$, after which the particle may bind an arbitrary number of times to various targets before reaching $\x$ at time $t$. Turning to the renewal equation (\ref{2Dren}b) for the absorption flux $\calJ_j(x_0,t)$, the
$j$th term on the first line represents all sample paths that are first adsorbed by the $j$th target in the time interval $[\tau,\tau+d\tau]$ and are subsequently absorbed at time $t$ without desorbing, which occurs with probability $(1-{\sigma}_j){\phi}_j(t-\tau)[\int_{\partial \calU_j}J_j(\y,\tau|\x_0)d\y]d\tau $. In a complementary fashion, the second term sums over all sample paths that are first adsorbed in the time interval $[\tau',\tau'+d\tau']$ at an arbitrary target $k$, desorb in the time interval $[\tau,\tau+d\tau]$ and are ultimately absorbed by the $j$th target at time $t$ following an arbitrary number of additional adsorption events.

Laplace transforming the renewal equations and using the convolution theorem yields
\begin{subequations}
\label{2DLT0}
 \begin{align}
 \wrho(\x,s|\x_0)&=\p(\x,s|\x_0) + \sum_{k=1}^N\sigma_k \wphi_k(s)\int_{\partial \calU_k} \wrho_k(\x,s|\y) \J_k(\y,s|\x_0)d\y, \\
\widetilde{\calJ}_j(\x_0,s)&= (1-{\sigma}_j)\wphi_j(s)\int_{\partial \calU_j}\J_j(\y,s|\x_0) d\y\nonumber \\
&\quad + \ \sum_{k=1}^N\sigma_k \wphi_k(s) \int_{\partial \calU_k}  \widetilde{\calJ}_j(\y,s) \J_k(\y,s|\x_0)d\y. \end{align}
 \end{subequations}
In order to solve these equations we still need to deal with the spatial surface integrals. In the case of a single target one could use a well known result from the classical theory of partial differential equations, namely, the solution of a general Robin BVP on a compact surface $\partial \calU$ can be computed in terms of the spectrum of a D-to-N (Dirichlet-to-Neumann) operator \cite{Grebenkov19a,Grebenkov23,Bressloff25}. However, the extension to multiple targets is generally not analytically tractable without some additional assumptions.

\subsection{Non-Markovian desorption and return to home base}
 
One scenario where we can eliminate the spatial integrals in the renewal equations is to assume that whenever the particle desorbs it immediately returns to its initial position $\x_0$ before continuing the diffusive search process.
In the case of a foraging animal, this could represent returning to its home base after failing to gather resources. For simplicity, we assume that the return time is much smaller than other characteristic times of the process so we can treat it as instantaneous. We can then exploit the fact that the renewal equations sew together successive rounds of adsorption/desorption. That is, we obtain the modified renewal equations
\begin{subequations}
\label{2DrenHB}
\begin{align}
 \rho(\x,t|\x_0)&=p(\x,t|\x_0)\\
 &+  \int_0^td\tau' \int_{\tau'}^t d\tau\,\rho(\x,t-\tau|\x_0) \sum_{k=1}^N  \sigma_k\phi_k(\tau-\tau') J_k(\x_0,\tau')\bigg ],\nonumber \\ 
 \calJ_j(\x_0,t)=&(1-{\sigma}_j) \int_0^td\tau {\phi}_j(t-\tau)      J_j(\x_0,\tau) \\
  &\quad  +   \int_0^td\tau' \int_{\tau'}^t d\tau\, \calJ_j(\x_0,t-\tau)\sum_{k=1}^N \sigma_k\phi_k(\tau-\tau')J_k(\x_0,\tau').\nonumber
 \end{align}
 \end{subequations}
Laplace transforming equations (\ref{2DrenHB}) gives
 \begin{subequations}
\label{2DLT0HB}
 \begin{align}
 \wrho(\x,s|\x_0)&=\p(\x,s|\x_0) +\wrho(\x,s|\x_0)\sum_{k=1}^N\sigma_k \wphi_k(s)  \J_k(\x_0,s), \\
\widetilde{\calJ}_j(\x_0,s)&=(1-{\sigma}_j) \wphi_j(s) \J_j(\x_0,s) +  \widetilde{\calJ}_j(\x_0,s)\sum_{k=1}^N\sigma_k \wphi_k(s)  \J_k(\x_0,s). \end{align}
 \end{subequations}
In particular, rearranging equation (\ref{2DLT0HB}b), shows that
 \begin{equation}
 \widetilde{\calJ}_j(\x_0,s)=\frac{(1-{\sigma}_j) \wphi_j(s) \J_j(\x_0,s) }{1-\sum_{k=1}^N\sigma_k \wphi_k(s)  \J_k(\x_0,s)}.
 \label{calJren}
 \end{equation}
 Hence, in this example, the absorption flux $\widetilde{\calJ}_j(\x_0,s)$ can be explicitly related to the corresponding adsorption fluxes 
  \begin{equation}
  \label{flu}
 \J_j(\x_0,s)=\kappa_0 \int_{\partial \calU_j}\p(\y,s|\x_0)d\y ,
 \end{equation}
with
  \begin{subequations}
 \label{LTRobin2}
\begin{align}
&D{\bm \nabla}^2 \p(\x,s|\x_0)-s\p(\x,s|\x_0)=-\delta(\x-\x_0),\quad \x \in \Omega \backslash \calU_a, \\
 &D{\bm \nabla}\p(\y,s|\x_0)\cdot \n_j(\y)=\kappa_0\p(\y,s|\x_0),\ \y\in \partial \calU_j.
 \end{align}
\end{subequations}

\section{Singular diffusion limit and matched asymptotics}

In \S 2 we considered three different models of the diffusive search for partially accessible targets and derived corresponding  equations for the Laplace transformed target fluxes. In the Markovian case, the fluxes are given by equation (\ref{MaJ}) and the solution of the full Robin BVP (\ref{LTRobin}). On the other hand, for non-Markovian desorption/absorption and return to home base, the fluxes are given by equation (\ref{calJren}) and the solution of the Robin BVP (\ref{LTRobin}) for irreversible adsorption. The third model assumes non-Markovian desorption in which the search process continues from the point of desorption. The target fluxes now satisfy the non-trivial integral equations (\ref{2DLT0}b). In this section we analyze the Robin BVPs and the latter integral equations in the small target limit where $\Omega$ can be treated as a singularly perturbed domain. For concreteness, suppose that each compartment is circularly symmetric. Denoting the radius and centre of the $j$-th compartment by $r_j$ and $\x_j$, respectively, we have
\begin{equation}
\calU_j=\{\x\in \Omega,\ |\x-\x_j|< r_j\},\quad \partial \calU_j=\{\x\in \Omega,\ |\x-\x_j|= r_j\}.
\end{equation}
The main characteristic of a singularly perturbed domain is that the targets $\calU_j$ are small compared to the size of the domain $\Omega$ and are well separated. More precisely, suppose that the domain $\Omega$ is inscribed by a rectangular area whose smallest dimension is $L$, and introduce the dimensionless parameter
$\epsilon =r_{\max}/L$ where $r_{\max}=\max \{r_j,\, j=1,\ldots,N\}$. We fix length scales by setting $L=1$ and writing $r_j=\epsilon \ell_j$ with $\ell_j=r_j/r_{\max}$,
and $0<\epsilon \ll 1$. We also assume that $|\x_i-\x_j| =O(1)$ for all $j\neq i$ and $\min_{{\bf s}}\{|\x_j -{\bf s}|,{\bf s} \in \partial \Omega \} =O(1)$, $j=1,\ldots,N$. (If $\Omega$ were unbounded, then we would identify $L$ with the smallest distance between any pair of compartments, that is, $L=\min\{|\x_i-\x_j|,\ i\neq j\}$.) In this section we first solve the 
Robin BVP (\ref{LTRobin2}) for irreversible adsorption using matched asymptotic analysis along the lines of Ref. \cite{Bressloff21A}. We then indicate how to modify the analysis in the case of the Robin BVP (\ref{LTRobin}) with desorption/absorption. Finally, we use the small-target approximation to solve equation (\ref{2DLT0}b) by eliminating the surface integrals.

\begin{figure}[b!]
\centering
\includegraphics[width=12cm]{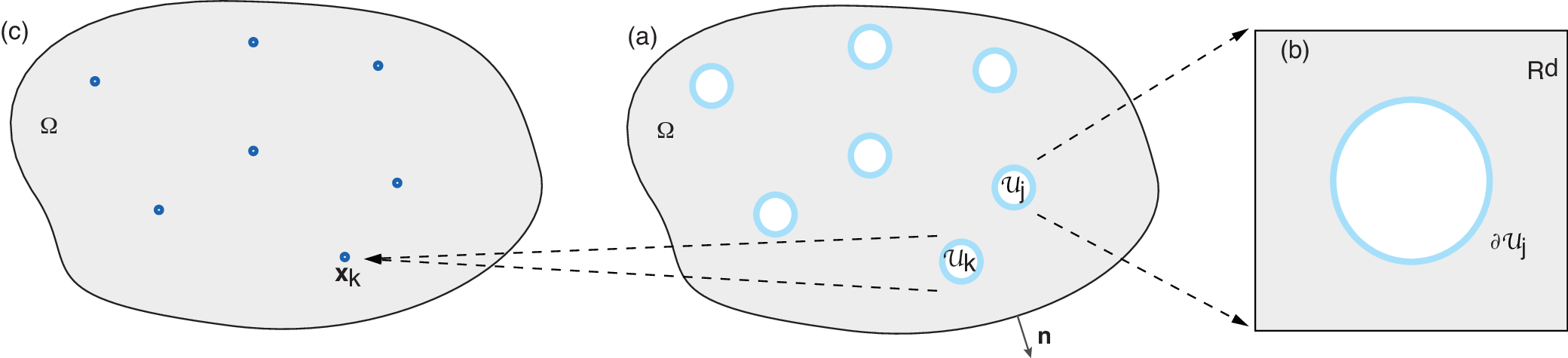} 
\caption{Formulation of the multi-target search process with irreversible adsorption as a singularly perturbed narrow capture problem. (a) Original unscaled domain. (b) Construction of the inner solution in terms of stretched coordinates $\z=\epsilon^{-1}(\x-{\x}_j)$, where ${\x}_j$ is the centre of the $j$-th target. The rescaled radius is $\ell_j$ and the region outside the compartment is taken to be $\R^2$ rather than the bounded domain $\Omega$. (c) Construction of the outer solution. The $k$th target is shrunk to a single point $\x_k$. The outer solution is expressed in terms of the corresponding modified Neumann Green's function and then matched with the inner solution around each target.}
\label{fig2}
\end{figure}

\subsection{Robin BVP for irreversible adsorption}
Following along analogous lines to previous studies of the Laplace transformed diffusion equation in 2D \cite{Lindsay16,Bressloff21A} and 3D \cite{Bressloff21B,Bressloff22}, we solve equations (\ref{LTRobin2}) in the small target limit using a combination of matched asymptotic analysis and Green's function methods. The basic idea is to construct an inner or local solution of the diffusion equation in a small neighbourhood of each compartment, which is then matched to an outer solution in the bulk domain, as illustrated in Fig. \ref{fig2}. The matching is achieved by expressing the outer solution in terms of the Green's function of the Laplace transformed diffusion equation in the absence of any targets. 
In the inner region around the $j$-th target, we introduce the stretched coordinates $\z=(\x-\x_j)/\epsilon$ and set $\p(\x_j+\epsilon \z,s|\x_0)=P_j(\z,s|\x_0)$ with
\begin{align}
D{\bm \nabla}_{\z}^2 P_j(\z,s|\x_0) =\epsilon^2 sP_j(\z,s|\x_0),\ |\z|>\ell_j,\quad  {\bm \nabla}_{\z} P_j\cdot \n_j=\epsilon \kappa_0 \calP_j,\quad |\z |=\ell_j .
\end{align}
Performing the rescaling $\kappa_0=\kappa_0'/\epsilon$ and dropping the $O(\epsilon^2)$ term yields an inner equation whose solution takes the form
\begin{equation}
P_j(\z,s|\x_0)\sim A_j(\x_0,s)\bigg [\ln (|\z|/\ell_j)+\frac{D}{\kappa_0'\ell_j}\bigg ].
\label{inner}
\end{equation}
The outer solution is constructed by shrinking each target to a single point and imposing a corresponding singularity condition as $\x\rightarrow \x_j$. The leading order contribution to the outer solution thus satisfies
\begin{subequations} 
\label{outer}
\begin{align}
	D{\bm \nabla}^2 \p_0(\x,s|\x_0) -s\p_0(\x,s|\x_0)&= -\delta(\x-\x_0) , \ \x\in \Omega',\quad
	{\bm  \nabla} \p_0\cdot \n=0, \ \x\in \partial \Omega,	\end{align}
	 with $\Omega':= \Omega\backslash \{\x_1,\ldots,\x_N\}$ and
	 \begin{equation}
	 \p_0(\x,s|\x_0)\sim  A_j(\x_0,s) \ln(|\x-\x_j|/\epsilon)  \quad \mbox{as} \ \x\rightarrow \x_j.
	 \end{equation}
\end{subequations}
In other words, we have the following equation for $\p_0$ on $\Omega$:
\begin{subequations}
 \begin{align}
	D\nabla^2 \p_0(\x,s|\x_0) -s\p_0(\x,s|\x_0)&= 2\pi D \sum_{j=1}^N A_j(\x_0,s)\delta(\x-\x_j)-\delta(\x-\x_0),\ \x \in \Omega , \\
	 \nabla \p_0 \cdot \n&=0, \ \x\in \partial \Omega.	\end{align}
	 \end{subequations}
Applying the divergence theorem to this equation implies that
\begin{equation}
\label{cond}
2\pi  D\sum_{j=1}^NA_j(\x_0,s)=1-s\int_{\Omega}\p_0(\x,s|\x_0)d\x.
\end{equation}
The outer equation then has the solution
\begin{equation}
\p_0(\x,s|\x_0)=G(\x,s|\x_0)-2\pi D \sum_{j=1}^N A_j(\x_0,s)G(\x,s|\x_j),
\end{equation}
where $G(\x,s|\x_0)$ is the Green's function of the modified Helmholtz equation:
\begin{subequations}
\label{GMH}
\begin{align}
	D{\bm \nabla}^2 G(\x,s|\x_0) -sG(\x,s|\x_0) &=-\delta(\x-\x_0) , \ \x\in \Omega,\\
	 {\bm \nabla} G(\x,s|x_0)\cdot \n  &=0,\ \x \in \partial \Omega, \\
	 G(\x,s|\x_0)&=-\frac{1}{2\pi D}\ln|\x-\x_0|+R(\x,s|\x_0),
	\end{align}
	\end{subequations}
and $R$ is the regular part of $G$.

We have $N$ unknown coefficients $A_j(\x_0,s)$, $j=1,\ldots,N$, which are obtained by solving $N$ constraints. The latter are constructed by matching the far-field behavior of the inner solutions $P_j$ with the near-field behavior of the outer solution $\p_0$ in a neighborhood of $\calU_k$ for $k=1,\ldots,N$:
\begin{align}
G(\x_k,s|\x_0)&=2\pi D\sum_{j\neq k}  A_{j}(\x_0,s)G(\x_k,s|\x_j)+A_k(\x_0,s)\bigg [\frac{1}{\nu}-\ln \ell_k+\frac{D}{\kappa_0' \ell_k}\bigg ]\nonumber \\
&\quad +2\pi  D A_{k}(\x_0,s)R(\x_k,s|\x_k),\qquad \nu=-\frac{1}{\ln \epsilon}.
\label{match1}
\end{align}
Equation (\ref{match1}) can be rewritten more compactly as the matrix equation
\begin{align}
\sum_{j=1}^N[{\bf I}+2\pi \nu D{\bm {\mathcal G}}(s)]_{kj}A_{j}(\x_0,s)=\nu G(\x_k,s|\x_0),
\end{align}
where
\begin{equation}
\label{calG}
{\mathcal G}_{jk}(s)= G(\x_j,s|\x_k),\ \mbox{for } i\neq j;\quad {\mathcal G}_{jj}(s)= R(\x_j,s|\x_j)-\frac{\dis \ln \ell_j}{\dis 2\pi D}+\frac{\dis 1}{\dis 2\pi \kappa_0'\ell_j}.
\end{equation}
We thus have the formal solution
\begin{equation}
\label{A}
{\bf A}(\x_0,s)=\nu [{\bf I}+2\pi \nu D{\bm {\mathcal G}}(s)]^{-1}{\bf G}(\x_0,s).
\end{equation}
We have introduced the vector ${\bf G}(\x_0,s)$ with components $ G_k(\x_0,s)=G(\x_k,s|\x_0)$.
The non-perturbative solution (\ref{A}) for the coefficient $A_j(\x_0,s)$ effectively sums over all logarithmic terms, which is equivalent to calculating the asymptotic solution for all terms of $O(\nu^k)$ for any $k$. This type of summation was originally obtained by Ward and Keller \cite{Ward93}, and is a common feature of strongly localized perturbations in 2D domains. Since $\nu \rightarrow 0$ more slowly than $\epsilon\rightarrow 0$, the summation over logarithmic terms yields $O(1)$ accuracy with respect to $\epsilon$.
Finally, the corresponding adsorption fluxes are obtained by substituting the inner solution into the Laplace transform of the flux equation (\ref{flu}):
the adsorption flux densities are 
\begin{align}
\J_k(\x_0,s)&=\kappa_0 \int_{\partial \calU_k} P_k((\y-\x_k)/\epsilon,s|\x_0) d\y=\kappa_0 |\partial \calU_k|\frac{DA_k(\x_0,s) }{\kappa_0' \ell_k}\nonumber \\
&=  2\pi DA_k(\x_0,s).
\label{moog}
\end{align}
Summing both sides of equation (\ref{moog}) with respect to $k$ and using equation (\ref{cond}) yields
\begin{equation}
\sum_{k=1}^N \widetilde{J}_k(\x_0,s)\sim 1-s\int_{\Omega}\wrho_0(\x,s|\x_0)d\x.
\end{equation}
This is a classical result relating the unconditional FPT density for adsorption to the survival probability in Laplace space.

 \subsection{Generalized Robin BVP for Markovian desorption}
 
 Comparison of equations (\ref{LTRobin}) and (\ref{LTRobin2}) shows that, in Laplace space, the only change needed to implement the generalized Robin boundary condition on $\partial \calU_k$ is to take $\kappa_0\rightarrow \kappa_0 g_k(s)$ with $g_k(s)$ given by equation (\ref{gs}). Otherwise the analysis is almost identical so we simply quote the results.
 First, the inner solution becomes
\begin{equation}
\calP_j(\z,s|\x_0)\sim \calA_j(\x_0,s)\bigg [\ln (|\z|/\ell_j)+\frac{D}{\kappa_0'g_j(s)\ell_j}\bigg ].
\label{inner2}
\end{equation}
Second the coefficients $ \calA_j(\x_0,s)$ satisfy the modified matrix equation
\begin{align}
G(\x_k,s|\x_0)&=2\pi D\sum_{j\neq k}  \calA_{j}(\x_0,s)G(\x_k,s|\x_j)+\calA_k(\x_0,s)\bigg [\frac{1}{\nu}-\ln \ell_k+\frac{D}{\kappa_0'g_k(s)\ell_k}\bigg ]\nonumber \\
&\quad +2\pi  D \calA_{k}(\x_0,s)R(\x_k,s|\x_k).
\label{match2}
\end{align}
In order to highlight the effects of desorption, we rewrite (\ref{match2}) as the matrix equation
\begin{align}
\sum_{j=1}^N[{\bf I}+2\pi \nu D{\bm {\mathcal G}}(s)]_{kj}\calA_{j}(\x_0,s)+\nu \Gamma_k(s)\calA_{k}(\x_0,s) =\nu G(\x_k,s|\x_0),
\end{align}
where
\begin{equation}
 \Gamma_k(s) =\frac{\dis D}{\dis \kappa_0' \ell_k}\bigg [\frac{\gamma_k}{s+\overline{\gamma}_k}\bigg ] .
\end{equation}
Introducing the matrix ${\bm \Theta}(s)$ with
\begin{equation}
\Theta_{jk}(s)=[{\bf I}+2\pi \nu D{\bm {\mathcal G}}(s)]_{jk}^{-1},
\end{equation}
we can expand $\calA_j$ as the Neumann series
\begin{align}
\label{Ajax}
\calA_j(\x_0,s)&=A_j(\x_0,s)-\nu \sum_{k=1}^N\Theta_{jk}(s)\Gamma_k(s)A_k(\x_0,s)\\
&\quad +\nu^2 \sum_{k,l=1}^N\Theta_{jl}(s)\Gamma_l(k)\Theta_{lk}(s)\Gamma_k(s)A_k(\x_0,s)\ldots  ,\nonumber 
\end{align}
where $A_k(\x_0,s)$ is the corresponding amplitude for irreversible adsorption, see equation (\ref{A}). The $n$-th term in the expansion represents the contribution from $n-1$ rounds of desorption and adsorption prior to absorption.
The right-hand side can be formally summed to give
\begin{equation}
\label{calA}
{\bm \calA}(\x_0,s)=\left [{\bf I}+\nu {\bm \calM}(s)\right ]^{-1}{\bf A}(\x_0,s),
\end{equation}
where
\begin{equation}
{\bm \calM}(s)={\bm \Theta}(s)\diag (\Gamma_1(s),\ldots,\Gamma_N(s)).
\end{equation}
Finally, the corresponding absorption fluxes are obtained by substituting the inner solution into the Laplace transform of the flux equation (\ref{abJk}):
\begin{align}
\label{abJk2}
\widetilde{\calJ}_k(\x_0,s)&\sim\frac{\kappa_0 \overline{\gamma}_k}{s+\gamma_k+\overline{\gamma}_k} \int_{\partial \calU_k} \calP_k((\y-\x_k)/\epsilon,s|\x_0)d\y\nonumber \\
&=\frac{\kappa_0 \overline{\gamma}_k}{s+\gamma_k+\overline{\gamma}_k} |\partial \calU_k| \frac{D\calA_k(\x_0,s) }{\kappa_0' g_k(s)\ell_j}=\frac{2\pi  D\overline{\gamma}_k}{s+\overline{\gamma}_k}\calA_k(\x_0,s).  
\end{align}

\subsection{Renewal equations for non-Markovian desorption and continuous search}

We end this section by showing how the small-target limit can be used to eliminate the spatial integrals in the renewal equation (\ref{2DLT0}b), and that this recovers equation (\ref{Ajax}) in the Markovian case. First, we substitute the inner solution (\ref{moog}) for irreversible adsorption into the right-hand side of (\ref{2DLT0}b) to give
 \begin{align}
\widetilde{\calJ}_j(\x_0,s)&\sim 2\pi D(1-{\sigma}_j)\wphi_j(s) A_j(\x_0,s)  \nonumber \\
&\quad + 2\pi D \sum_{k=1}^N\sigma_k \wphi_k(s) A_k(\x_0,s)\frac{1}{|\partial \calU_k|}\int_{\partial \calU_k}  \widetilde{\calJ}_j(\y,s) d\y. \label{step0}\end{align}
Second, we set $\x_0=\y'\in \partial \calU_l$ and average with respect to $\y'$. This yields the matrix equation
 \begin{align}
\widetilde{\calJ}_j(\bx_l,s)&\sim2\pi D (1-\sigma_j)\wphi_j(s)A_j(\bx_l,s) +2\pi D\sum_{k=1}^N \sigma_k\wphi_k(s)  \widetilde{\calJ}_j(\bx_k,s) A_k(\bx_l,s),
\label{step}
\end{align}
with
\begin{equation}
A_k(\bx_l,s)=\frac{1}{|\partial \calU_l|}\int_{\partial \calU_l}A_k(\y',s)d\y'
\end{equation}
etc. Iterating equation (\ref{step}) and substituting into (\ref{step0}), we obtain the Neumann series expansion
\begin{align}
\frac{\widetilde{\calJ}_j(\x_0,s)}{2\pi D}&=  (1-{\sigma}_j)\wphi_j(s) \bigg [A_j(\x_0,s) + 2\pi D \sum_{k=1}^N\sigma_k \wphi_k(s) A_k(\x_0,s)A_j(\bx_k,s)
 \nonumber \\
&\quad + (2\pi D)^2 \sum_{k,l=1}^N\sigma_k \wphi_k(s) \sigma_l\wphi_l(s)A_k(\x_0,s)A_l(\bx_k,s)A_j(\bx_l,s)+\ldots \big ].
\label{mono}
\end{align}
Next, from equation (\ref{A}) we have
\begin{equation}
\label{matrix2}
A_{j}(\bx_l,s)= \nu \sum_{k=1}^N\Theta_{jk}(s)G(\x_k,s|\bx_l) ,
\end{equation}
and
\begin{equation}
G(\x_k,s|\bx_l):=\frac{1}{|\partial \calU_l|}\int_{\partial \calU_l}G(\x_k,s|\y')d\y',\quad \y \in \partial \calU_l .
 \end{equation}
 If $k\neq l$ then we can take $G(\x_k,s|\bx_l)\approx G(\x_k,s|\x_l)$ since the targets are well-separated, whereas
 \begin{align}
 G(\x_l,s|\bx_l) &= \frac{1}{|\partial \calU_l|}\int_{\partial \calU_l} \bigg [-\frac{1}{2\pi D}\ln |\y-\x_l|+R(\x_l,s|\y)\bigg ]d\y\nonumber \\
 & \approx R(\x_l,s|\x_l)-\frac{\dis \ln \ell_l}{\dis 2\pi D}+\frac{1}{2\pi \nu D}.
 \end{align}
Hence, equation (\ref{matrix2}) implies that
\begin{align}
2\pi DA_{j}(\bx_l,s)&=  \delta_{j,l}  -\frac{\dis \nu D}{\dis \kappa_0'\ell_l}\Theta_{jl}(s) .
\label{man}
\end{align}
 Plugging the last result into the Neumann series (\ref{mono}) up to the cubic term in the amplitudes $A_k$ gives
\begin{align}
\frac{\widetilde{\calJ}_j(\x_0,s)}{2\pi D}&=  (1-{\sigma}_j)\wphi_j(s) \bigg [A_j(\x_0,s) +\sigma_j\wphi_j(s) A_j(\x_0,s) +  [\sigma_j \wphi_j(s)]^2 A_j(\x_0,s)+\ldots
\nonumber \\
&\quad -(1+\sigma_j\wphi_j(s)+\ldots)\sum_{k=1}^N\Theta_{jk}(s)\bigg (\sigma_k \wphi_k(s)\frac{\dis \nu D}{\dis \kappa_0'\ell_k}\bigg )A_k(\x_0,s)
 \nonumber \\
&\quad + \sum_{k=1}^N\Theta_{jk}(s)\bigg ([\sigma_k \wphi_k(s)]^2\frac{\dis \nu D}{\dis \kappa_0'\ell_k}\bigg )A_k(\x_0,s)+\ldots 
\label{mono3}\\
&\quad + \sum_{k,l=1}^N\Theta_{jl}(s)\bigg (\sigma_l\wphi_l(s)\frac{\dis \nu D}{\dis \kappa_0'\ell_l}\bigg )\Theta_{lk}(s)\bigg (\sigma_k \wphi_k(s)\frac{\dis \nu D}{\dis \kappa_0'\ell_k}\bigg )A_k(\x_0,s)+\ldots  \bigg ] .\nonumber
\end{align}
Including higher terms establishes a set of geometric series that can be summed to yield the effective Neumann series
\begin{align}
\label{main1}
&\frac{\widetilde{\calJ}_j(\x_0,s)}{2\pi D} \\
&\sim \frac{(1-{\sigma}_j)\wphi_j(s) }{1-\sigma_j\wphi_j(s)}\bigg [A_j(\x_0,s) 
 -\nu \sum_{k=1}^N\Theta_{jk}(s)\bigg (\frac{\sigma_k \wphi_k(s)}{1-\sigma_k\wphi_k(s)}\frac{\dis  D}{\dis \kappa_0'\ell_k}\bigg )A_k(\x_0,s)
 \nonumber \\
&\quad + \nu^2\sum_{k,l=1}^N\Theta_{jl}(s)\bigg (\frac{\sigma_l\wphi_l(s)}{1-\sigma_l\wphi_l(s)}\frac{\dis  D}{\dis \kappa_0'\ell_l}\bigg )\Theta_{lk}(s)\bigg (\frac{\sigma_k \wphi_k(s)}{1-\sigma_k\wphi_k(s)}\frac{\dis  D}{\dis \kappa_0'\ell_k}\bigg )A_k(\x_0,s)+\ldots \bigg ]  .\nonumber
\end{align}
Analogous to equation (\ref{Ajax}), the $n$th term in the expansion represents the effective contribution from $n-1$ rounds of desorption and adsorption. Moreover, the Neumann series can be formally summed to yield the non-perturbative result
\begin{equation}
\label{calA2}
\frac{\widetilde{\calJ}_j(\x_0,s)}{2\pi D}=\frac{(1-{\sigma}_j)\wphi_j(s) }{1-\sigma_j\wphi_j(s)}\left [{\bf I}+\nu \widehat{\bm \calM}(s)\right ]^{-1}{\bf A}(\x_0,s),
\end{equation}
where
\begin{equation}
\widehat{\bm \calM}(s)={\bm \Theta}(s)\diag (\widehat{\Gamma}_1(s),\ldots,\widehat{\Gamma}_N(s)),
\end{equation}
and
\begin{equation}
\widehat{\Gamma}_k(s)=\frac{\sigma_k \wphi_k(s)}{1-\sigma_k\wphi_k(s)}\frac{\dis  D}{\dis \kappa_0'\ell_k}.
\end{equation}
Finally, we can check that equation (\ref{main1}) recovers equation (\ref{Ajax}) in the Markovian case (\ref{ephi}). More specifically,
\begin{equation}
\frac{\sigma_k \wphi_k(s)}{1-\sigma_k\wphi_k(s)}=\frac{\gamma_k}{s+\overline{\gamma}_k},\quad \frac{(1-{\sigma}_j)\wphi_j(s) }{1-\sigma_j\wphi_j(s)}=\frac{\overline{\gamma}_j}{s+\overline{\gamma}_j}.
\end{equation}

\section{Splitting probabilities and MFPTs for absorption}

We now use our asymptotic results for the target fluxes $\widetilde{\calJ}_k(\x_0,s)$ to calculate the splitting probabilities and MFPTs for absorption. The first step is to perform a small-$s$ expansion of the amplitudes $A_j(\x_0,s)$ satisfying equation (\ref{match1})\footnote{Formally solving equation (\ref{match1}) to obtain the matrix solutions (\ref{A}) and (\ref{matrix2}) is only valid when $s>0$ so we cannot apply the limit $s\rightarrow 0$ directly to equations (\ref{A}) and (\ref{matrix2}). This is due to the fact that the Neumann Green's function is singular in the limit $s\rightarrow 0$ and care has to be taken in cancelling these singularities. }. The coefficients in this expansion represent the splitting probabilities and FPT moments in the case of irreversible adsorption. We then substitute the small-$s$ expansion into equations (\ref{calJren}) and (\ref{mono}) to extract the corresponding results for absorption with and without a return to home base, respectively.

 \subsection{Small-$s$ expansion}

Following along analogous lines to Ref. \cite{Bressloff21A}, we perform a small-$s$ expansion of the coefficients $A_k(\x_0,s)$ by expanding equation (\ref{match1}) using the following $s$-dependence of the Green's function $G$:
\begin{equation}
\label{Gs}
G(\x,s|\x_0)=\frac{1}{s|\Omega|}+G_0(\x,\x_0)+sG_1(\x,\x_0)+o(s),
\end{equation}
where $G_0$ is the modified Neumann Green's function of the steady-state diffusion equation:
\begin{subequations}
\label{GM}
\begin{align}
	D\nabla^2 G_0(\x,\x_0)  &=\frac{1}{|\Omega|}-\delta(\x-\x_0) , \ \x\in \Omega,\\
	 \nabla G_0(\x,\x_0)\cdot \n  &=0,\ \x \in \partial \Omega,\quad \int_{\Omega}G_0(\x,\x_0)d\x=0, \\
	 G_0(\x,\x_0)&=-\frac{1}{2\pi}\ln|\x-\x_0|+R_0(\x,\x_0),
	\end{align}
	\end{subequations}
	and
	\begin{subequations}
\label{GM1}
\begin{align}
	D\nabla^2 G_1(\x,\x_0)  -G_0(\x,\x_0)&=0 , \ \x\in \Omega,\\
	 \nabla G_1(\x,\x_0)\cdot \n  &=0,\ \x \in \partial \Omega,\quad \int_{\Omega}G_1(\x,\x_0)d\x=0.	\end{align}
	\end{subequations}
	Note that the solution of equation (\ref{GM1}) is non-singular with
	\begin{equation}
	G_1(\x,\x_0)=-\int_{\Omega}G_0(\y,\x)G_0(\y,\x_0)d\y.
	\end{equation}
Taking the limit $s\rightarrow 0$ in equation (\ref{cond}) implies that
\begin{equation}
2\pi  D \lim_{s\rightarrow 0}\sum_{j=1}^NA_j(\x_0,s)=1-\lim_{s\rightarrow 0}s\int_{\Omega}\p_0(\x,s|\x_0)d\x=1,
\end{equation}
since $\lim_{s\rightarrow 0}s\p_0(\x,s|\x_0)=\lim_{t\rightarrow \infty} p_0(\x,t|\x_0)=0$.
Hence, the coefficient $A_k(\x_0,s)$ has a small-$s$ expansion of the form
\begin{equation}
\label{anu}
A_k(\x_0,s)={A}^{(0)}_k(\x_0)+s\chi_k(\x_0)+O(s^2)
\end{equation}
with
\begin{equation}
\label{con0}
 \sum_{k=1}^N{A}^{(0)}_k(\x_0)=\frac{1}{2\pi D}.
\end{equation}

Substituting equations (\ref{Gs}) and (\ref{anu}) into the matching equation (\ref{match1}) and keeping terms up to $O(s)$ yields
\begin{align}
\label{Nexp}
&\frac{1}{s|\Omega|}+G_0(\x_k,\x_0)+sG_1(\x_k,\x_0)\\
&=2\pi   D\sum_{j\neq k}  \bigg [{A}^{(0)}_j(\x_0)+s  \chi_j(\x_0)\bigg ]\bigg [\frac{1}{s|\Omega|}+G_0(\x_k,\x_j)+sG_1(\x_k,\x_j)\bigg ]\nonumber \\
&+  \bigg [{A}^{(0)}_k(\x_0)+s\chi_k(\x_0)\bigg ]\bigg [\frac{1}{\nu}-\ln \ell_k+\frac{D}{\kappa_0' \ell_k}\bigg ]\nonumber \\
& +2\pi   D \bigg [{A}^{(0)}_k(\x_0)+s  \chi_k(\x_0)\bigg ]\bigg [\frac{1}{s|\Omega|}+R_0(\x_k,\x_k)+sR_1(\x_k,\x_k)\bigg ].\nonumber 
\end{align}
The $O(1/s)$ terms cancel due to the constraint (\ref{con0}). Collecting the $O(1)$ terms gives
\begin{align}
\label{match3}
&2\pi  D\sum_{j\neq k} {A}^{(0)}_j(\x_0) G_0(\x_k,\x_j)+2\pi D {A}^{(0)}_k(\x_0)R_0(\x_k,\x_k) \\
&\quad + {A}^{(0)}_k(\x_0) \bigg [\frac{1}{\nu}-\ln \ell_k+\frac{D}{\kappa_0' \ell_k} \bigg ]
=G_0(\x_k,\x_0)-\chi(\x_0),\nonumber 
\end{align}
where 
\begin{equation}
\chi(\x_0)=\frac{2\pi  D}{|\Omega|}\sum_{k=1}^N\chi_k(\x_0). 
\end{equation}
Introducing the matrix ${\bm \calG}^{(0)}$ with elements
\begin{equation}
\calG^{(0)}_{kj}= G_0(\x_k,\x_j),\ k\neq j,\quad \calG^{(0)}_{kk}=  R_0(\x_k,\x_k)+\frac{1}{2\pi D}\bigg [-\ln \ell_k+\frac{D}{\kappa_0' \ell_k}\bigg ],
\end{equation}
we can write the solution as
\begin{equation}
{A}^{(0)}_k(\x_0)=\nu \sum_{j=1}^N[{\bf I}+2\pi \nu D{\bm \calG}^{(0)}]^{-1}_{kj} \bigg (G_0(\x_j,\x_0)- \chi(\x_0)\bigg ).
\label{hier1}
\end{equation}
Finally, the unknown coefficient $\chi(\x_0)$ is determined by imposing the constraint in (\ref{con0}):
\begin{align}
\frac{1}{2\pi \nu D}=\sum_{j,k=1}^N[{\bf I}+2\pi \nu D {\bm \calG}^{(0)}]^{-1}_{kj} \bigg (G_0(\x_j,\x_0)- \chi(\x_0)\bigg ).
\label{hier0}
\end{align}

Equations (\ref{moog}), (\ref{anu}), (\ref{hier1}) and (\ref{hier0}) can now be used to determine the splitting probabilities and unconditional MFPT for irreversible adsorption: 
     \begin{align}
   \overline{\pi}_k(\x_0)&:= \lim_{s\rightarrow 0}\J_k(x_0,s) \sim 2\pi D {A}^{(0)}_k(\x_0),  
   \label{piri} \end{align}
   and
\begin{align}
T(\x_0)&:
= \sum_{k=1}^N\left . \left (-\frac{d}{ds}\right ) \J_k(x_0,s)\right |_{s=0}\sim -2\pi D\sum_{k=1}^N\chi_k(\x_0)=-|\Omega| \chi(\x_0).
\label{TP}
\end{align}
Note that these results are non-perturbative with respect to the small parameter $\nu$. 
In addition, the conditional MFPTs are given by
 \begin{align}
\overline{\pi}_k(\x_0)T_k(\x_0)&:=\left . \left (-\frac{d}{ds}\right ) \J_k(x_0,s)\right |_{s=0}\sim -2\pi D \chi_k(\x_0).
 \end{align}
However, in order to determine the individual coefficients $\chi_k(\x_0)$, it is necessary to collect the $O(s)$ terms in the small-$s$ expansion (\ref{Nexp}).

\subsection{Return to home base after desorption} Equation (\ref{calJren}) implies that we can express the splitting probabilities and FPT moments for absorption in terms of the corresponding quantities for irreversible adsorption. This was previously explored in the case of a single target \cite{Bressloff25,Bressloff25a}. First consider the splitting probabilities. Taking the limit $s\rightarrow 0$ in equation (\ref{calJren}) shows that
\begin{equation}
\pi_j(\x_0)\sim \frac{(1-{\sigma}_j) \overline{\pi}_j (\x_0)}{\sum_{k=1}^N(1-\sigma_k) \overline{\pi}_k(\x_0)}
\label{resbpi}
\end{equation}
with $\overline{\pi}_j(\x_0)$ given by equations (\ref{hier1}), (\ref{hier0}) and (\ref{piri}).
We have used the identity $\sum_{j=1}^N \overline{\pi}_j =1$.
It immediately follows that $\sum_{j=1}^N\pi_j=1$.
Note the non-perturbative expression on the right-hand side of equation (\ref{resbpi}) can be expanded as a power series in $\nu$ by using equation (\ref{piri2}). The leading order contribution to the splitting probability is $1/N$ weighted by the relative probability of absorption $(1-\sigma_j)(1-\overline{\sigma})$, where $\overline{\sigma}=\sum_{k=1}^N \sigma_k/N$.

Next, assuming that $\widetilde{\phi}(s)$ has finite moments, we can substitute the following expansions into equation (\ref{calJren}):
 \begin{subequations}
 \label{MFPT}
 \begin{align}
 \J_k(\x_0,s)&=\overline{\pi}_k(\x_0) \left [1-sT_k(\x_0)+\frac{s^2}{2}T_k^{(2)}(\x_0)+O(s^3) \right ],\\
 \wphi_k(s)&=1-s\langle \tau\rangle_k +\frac{s^2}{2} \langle \tau^2\rangle_k+O(s^3).
 \end{align}
 \end{subequations}
 Collecting $O(s)$ terms generates the conditional MFPT relation
\begin{align}
&\pi_j(\x_0) \calT_j(x_0)\\
&=\frac{\overline{\pi}_j(\x_0)(1-\sigma_j)}{\sum_{k=1}^N\overline{\pi}_k(\x_0)(1-\sigma_k)}\bigg [T_j(\x_0)+  \langle \tau \rangle_j+\frac{\sum_{k=1}^N\overline{\pi}_k\sigma_k(T_k(\x_0)+  \langle \tau \rangle_k)}{\sum_{k=1}^N\overline{\pi}_k(\x_0)(1-\sigma_k)}\bigg ].\nonumber 
 \end{align}
 Moreover, summing both sides with respect $j$ implies that the relationship for the unconditional MFPT is
 \begin{align}
 \calT(\x_0)&=\sum_{j=1}^N\pi_j(\x_0) \calT_j(x_0) 
 =\frac{T(\x_0)+\sum_{k=1}^N\overline{\pi}_k(\x_0) \langle \tau \rangle_k}{\sum_{k=1}^N\overline{\pi}_k(\x_0)(1-\sigma_k)},
 \label{calT}
 \end{align}
since $T(\x_0)=\sum_{k=1}^N\overline{\pi}_k(\x_0)T_k(\x_0) $. Again this result is non-perturbative with respect to $\nu$.

In practice, it is often useful to Taylor expand the non-perturbative solutions in powers of $\nu$. Setting 
\begin{equation}
 \chi(\x_0)\sim - \frac{1}{2\pi \nu DN}+\chi_0(\x_0)+O(\nu)
 \end{equation}
 and substituting into (\ref{hier0}) gives
 \begin{align}
 \chi_0(\x_0)=\frac{1}{N}\sum_{j=1}^NG_0(\x_j,\x_0)-\frac{1}{N}\sum_{j,k=1}^N\calG^{(0)}_{kj}.
 \end{align}
and
\begin{align}
{A}^{(0)}_k(\x_0)=\frac{1}{2\pi DN}+\nu \bigg [G_0(\x_k,\x_0)- \sum_{j=1}^N\calG^{(0)}_{kj}- \chi_0(\x_0)\bigg ]+O(\nu^2).
\label{AA}
\end{align}
We thus obtain the asymptotic expansions
 \begin{align}
 \label{piri2}
   \overline{\pi}_k(\x_0)   &\sim \frac{1}{ N}+2\pi \nu D\bigg [G_0(\x_k,\x_0)- \sum_{j=1}^N\calG^{(0)}_{kj}- \chi_0(\x_0)\bigg ]+O(\nu^2).
   \end{align}
   and
\begin{align}
T(\x_0)&\sim \frac{|\Omega|}{2\pi \nu ND }-\frac{|\Omega|}{N} \bigg [\sum_{j=1}^NG_0(\x_j,\x_0)- \sum_{j,k=1}^N\calG^{(0)}_{kj}\bigg ]+O(\nu).
\label{TP2}
\end{align}

\subsubsection{Pair of targets in the unit disc}

\begin{figure}[t!]
\centering
\includegraphics[width=8.5cm]{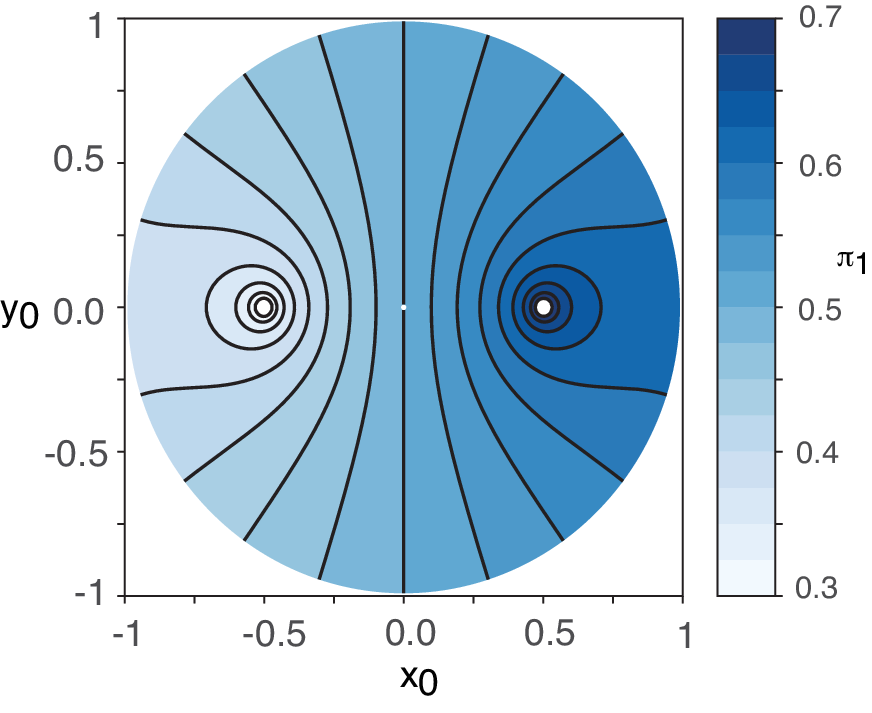} 
\caption{A pair of identical targets of radii $\epsilon$ are placed at the positions $\x_1=(a,0)$ and $\x_2=(-a,0)$ in the unit disc with $a=0.5$. The contour plot shows the splitting probability $\overline{\pi}_1(\x_0)$ for irreversible adsorption, see equation (\ref{2targ}a), as a function of $\x_0=(r\cos \theta,r\sin \theta)$ with $0<r<1$ and $0\leq \theta \leq 2\pi$. Other parameters are $\nu=0.1$, $D=1$ and $\kappa_0\rightarrow \infty$.}
\label{fig4}
\end{figure}

\begin{figure}[b!]
\centering
\includegraphics[width=13cm]{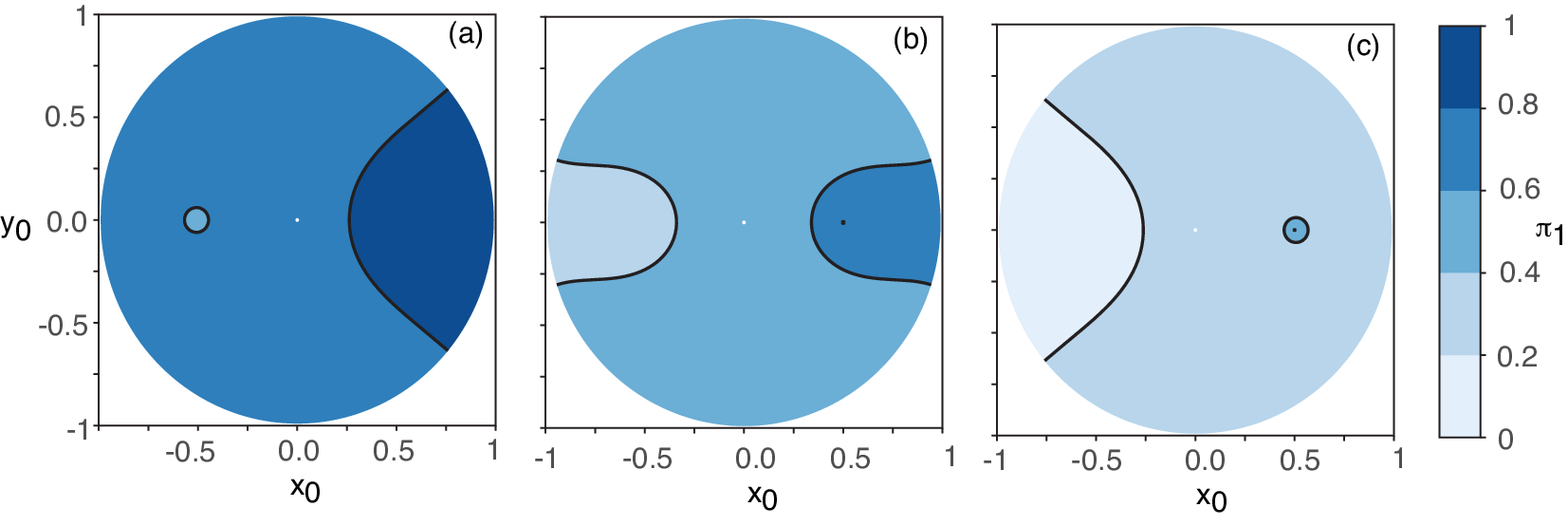} 
\caption{Contour plots of the splitting probability $\pi_1(\x_0)$ for absorption, equation (\ref{resbpi}), as a function of $\x_0=(r\cos \theta,r\sin \theta)$: (a) $\sigma_1=0.25,\sigma_2 =0.75$, (b) $\sigma_1=\sigma_2=0.5$, (c) $ \sigma_1 =0.75,\sigma_2=0.25$. Same target configuration and other parameters as Fig. \ref{fig4}.}
\label{fig5}
\end{figure}

\begin{figure}[t!]
\centering
\includegraphics[width=13cm]{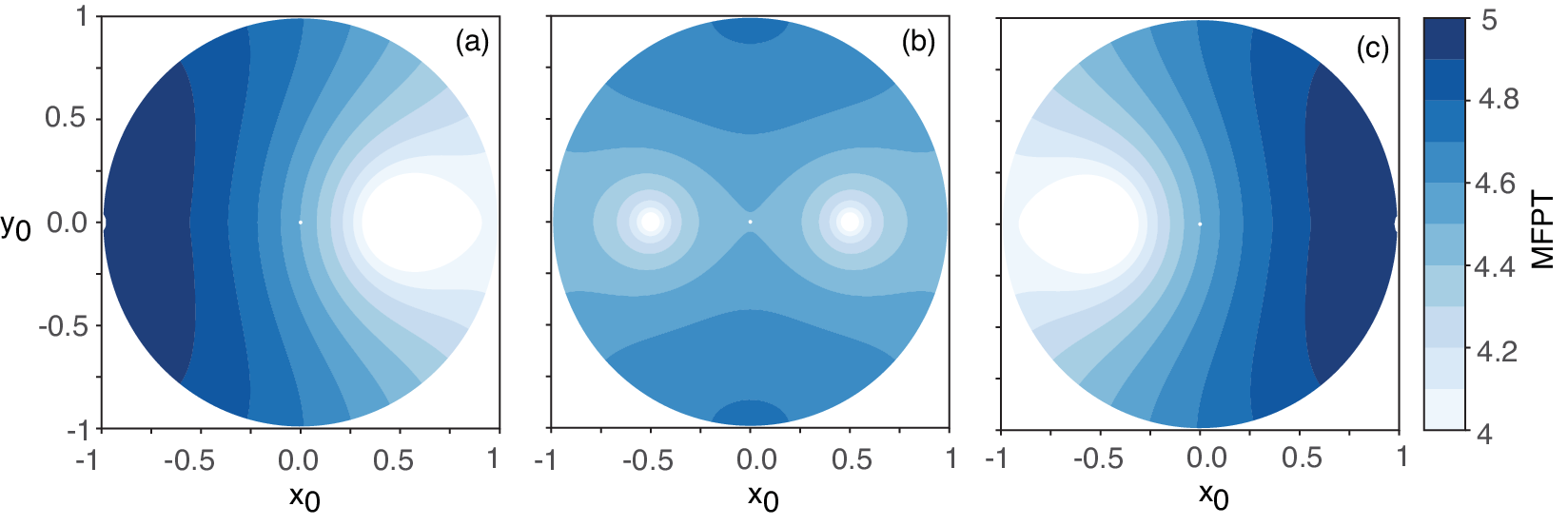} 
\caption{Contour plots of the unconditional MFPT $\calT(\x_0)$ for absorption, equation (\ref{calT}), as a function of $\x_0=(r\cos \theta,r\sin \theta)$:  (a) $\sigma_1=0.25,\sigma_2 =0.75$, (b) $\sigma_1=\sigma_2=0.5$, (c) $ \sigma_1 =0.75,\sigma_2=0.25$. We also set $\langle \tau\rangle =0$ for both targets.Same target configuration and other parameters as Fig. \ref{fig4}.}
\label{fig6}
\end{figure}

As a simple illustration of the above analysis, suppose that $\Omega$ is the unit disc containing a pair of identical targets placed on the $x$-axis at $\x_{1}=(a,0)$ and $\x_2=(-a,0)$, $0<a<1$. The Neumann Green's function $G_0(\x,\xi)$ in the unit disc is known explicitly:
\begin{align}
G_0(\x,\hxi)&=\frac{1}{2\pi}\left [-\ln(|\x-\hxi|)-\ln\left (\left |\x|\hxi|-\frac{\hxi}{|\hxi |}\right |\right )  +\frac{1}{2}(|\x|^2+|\hxi|^2)-\frac{3}{4}\right ],
\label{Gdisc}
\end{align}
with the regular part obtained by dropping the first logarithmic term. Assuming that $\ell_1=\ell_2=1$, $D=1$ and $\kappa_0'\gg 1$ (fast adsorption),
the splitting probabilities and unconditional MFPT for irreversible adsorption have the leading-order asymptotic expansions (see (\ref{piri2}) and (\ref{TP2}))
\begin{subequations}
\label{2targ}
\begin{align}
   \overline{\pi}_1(\x_0)   &\sim \frac{1}{ N}+2\pi \nu \bigg [G_0(\x_1,\x_0)-G_0(\x_1,\x_2)-R_0(\x_1,\x_1)- \chi_0(\x_0)\bigg ] ,\\
     \overline{\pi}_2(\x_0)   &\sim \frac{1}{ N}+2\pi \nu \bigg [G_0(\x_2,\x_0)-G_0(\x_2,\x_1)-R_0(\x_2,\x_2)- \chi_0(\x_0)\bigg ] ,\\
T(\x_0)&\sim \frac{|\Omega|}{4\pi \nu }-\frac{|\Omega|}{2} \chi_0(\x_0) ,
\end{align}
\end{subequations}
with $|\Omega|=\pi$ and
 \begin{align}
 \chi_0(\x_0)&=\frac{1}{2}\bigg [\ G_0(\x_1,\x_0)+G(\x_2,\x_0)-G(\x_1,\x_2)\nonumber \\
 &\quad -G(\x_2,\x_1) -R_0(\x_1,\x_1)-R_0(\x_2,\x_2)\bigg ].
 \end{align}
 In Fig. \ref{fig4} we show a contour plot of $ \overline{\pi}_1(\x_0)$ in equation (\ref{2targ}a)) as a function of the initial position $\x_0=(r\cos \theta,r\sin\theta)$. From the symmetry of the target configuration, we have $\overline{\pi}_1(r,-\theta) =\overline{\pi}_1(r,\theta) $ and $\overline{\pi}_1(r,\pi-\theta) =1-\overline{\pi}_1(r,\theta) $. Given $\overline{\pi}_1(r,\theta) $, the corresponding splitting probability $\pi_1(\x_0)$ for absorption is obtained from equation (\ref{resbpi}). The effect of the desorption probability $ \sigma_2$ on $\pi_1(\x_0)$ for fixed $\sigma_1=0.5$ is illustrated in Fig. \ref{fig5}. The latter shows contour plots of $\pi_1(\x_0)$ for the three cases $\sigma_2<\sigma_1$, $\sigma_2=\sigma_1$ and $\sigma_2>\sigma_1$. Clearly, increasing the relative probability of desorption for target 1 reduces $\pi_1(\x_0)$ for all $\x_0$. Analogous plots of the unconditional MFPT $\calT(\x_0)$ are shown in Fig. \ref{fig6}. For the sake of illustration, we assume that the mean waiting times between successive desorption/absorption events are negligible. In this case, increasing the relative probability of desorption for target 1 biases the region of smaller MFPTs to initial positions closer to target 2.

\subsection{Continuous search after desorption}

Extracting the splitting probabilities and MFPTs for absorption by substituting the small-$s$ expansions (\ref{MFPT}) into equation (\ref{mono}) is considerably more involved, and require some form of diagrammatic formulation to be effective. Therefore, here we restrict our analysis to the relatively simple case of the splitting probabilities.

Taking the limit $s\rightarrow 0$ of equation (\ref{mono}) yields the following series expansion of the splitting probabilities:
\begin{align}
\pi_j(\x_0)&=  (1-{\sigma}_j) \bigg [\overline{\pi}_j(\x_0) + \sum_{k=1}^N\sigma_k  \overline{\pi}_k(\x_0) \overline{\pi}_j(\bx_k)
 \nonumber \\
&\quad +   \sum_{k,l=1}^N\sigma_k  \sigma_l \overline{\pi}_k(\x_0)\overline{\pi}_l(\bx_k)\overline{\pi}_j(\bx_l)+\ldots \big ].
\label{mono2}
\end{align}
with
\begin{equation}
\overline{\pi}_j(\bx_l)=2\pi D{A}_j^{(0)}(\bx_l)=\frac{2\pi D}{|\partial \calU_l|}\int_{\partial \calU_l}{A}^{(0)}_j(\y')d\y'
\end{equation}
and ${A}^{(0)}(\y)$ given by equation (\ref{hier1}). Formally summing equation (\ref{mono2}) yields the non-perturbative result
\begin{equation}
\label{matte}
\pi_j(\x_0)\sim (1-\sigma_j)\sum_{k=1}^N [{\bf I}-{\bf Q}]_{jk}^{-1}\overline{\pi}_k(\x_0),
\end{equation}
with $Q_{jk}=\overline{\pi}_j(\overline{\x}_k)\sigma_k$.
Setting $\x_0=\y'\in \partial \calU_l$ in equations (\ref{hier1})--(\ref{hier0})  and averaging with respect to $\y'$ shows that
\begin{align}
{A}^{(0)}_k(\bx_l)&=\nu \sum_{j=1}^N[{\bf I}+2\pi \nu D{\bm \calG}^{(0)}]^{-1}_{kj} \bigg (G_0(\x_j,\bx_l)- \chi(\bx_l)\bigg ),
\label{zhier1}\\
\frac{1}{2\pi \nu D}&=\sum_{j,k=1}^N[{\bf I}+2\pi \nu D {\bm \calG}^{(0)}]^{-1}_{kj} \bigg (G_0(\x_j,\bx_l)- \chi(\bx_l)\bigg ),
\label{zhier0}
\end{align}
where
 \begin{align}
 G_0(\x_j,\bx_l)\approx G_0(\x_j,\x_l),\ j\neq l, \quad G_0(\x_l,\bx_l) 
  \approx R_0(\x_l,\x_l)-\frac{\dis \ln \ell_l}{\dis 2\pi D}+\frac{1}{2\pi \nu D}.
 \end{align}

We now obtain the leading asymptotic expansion of the splitting probabilities with respect to $\nu$. Performing an expansion of equation (\ref{zhier0}) in powers of $\nu$, we have 
\begin{align}
\frac{1}{2\pi \nu D}&\sim \sum_{j\neq l}G_0(\x_j,\x_l)+R_0(\x_l,\x_l)-\frac{\dis \ln \ell_l}{\dis 2\pi D}+\frac{1}{2\pi \nu D}-N \chi(\bx_l)-\sum_{k=1}^N {\bm \calG}^{(0)}_{kl} \nonumber \\
&\quad +O(\nu),
\end{align}
which can be rearranged to give $ \chi(\bx_l)\sim  \chi_0(\bx_l)+O(\nu)$ with
\begin{align}
 \chi_0(\bx_l)=\frac{1}{N}\bigg [ \sum_{j\neq l}G_0(\x_j,\x_l)+R_0(\x_l,\x_l)-\frac{\dis \ln \ell_l}{\dis 2\pi D}-\sum_{k=1}^N {\bm \calG}^{(0)}_{kl}\bigg ]=-\frac{\dis 1}{\dis 2\pi  N\kappa_0'\ell_l}.
 \end{align}
We have used equations (\ref{calG}).
It follows that
\begin{subequations}
\label{ipip}
\begin{align}
\overline{\pi}_l(\bx_l)  &\sim 1-\left (1-\frac{1}{N}\right )\frac{\dis \nu D}{\dis \kappa_0'\ell_l}, \\
  \overline{\pi}_k(\bx_l)  &\sim \frac{\dis \nu D}{\dis  N\kappa_0'\ell_l},\quad k\neq l.
\end{align}
\end{subequations}
 Hence, if the particle continues diffusing from the point of desorption on the $l$th target surface $\partial \calU_{\ell}$, then the probability that it is subsequently re-adsorbed by the same target is $O(1)$, whereas the probability that it rebinds to a different target is $O(\nu)$. This is consistent with the assumption that the targets are well separated compared to their sizes. Using equations (\ref{ipip}) we have the following asymptotic expansion of the matrix appearing in equation (\ref{matte}):
\begin{align}
[{\bf I}-{\bf Q}]_{jk}\sim \delta_{j,k}(1-\sigma_j)+\frac{\dis \nu D\sigma_j }{\dis  \kappa_0'\ell_j}\delta_{j,k}-\frac{\dis \nu D\sigma_k }{\dis  N\kappa_0'\ell_k}.
\end{align}
Hence, we can write
\begin{equation}
{\bf I}-{\bf Q}\sim {\bf D}({\bf I}-\nu {\bf E}),
\end{equation}
where
\begin{align}
D_{jl}=(1-\sigma_j)\delta_{j,l},\quad E_{lk}=-\frac{\dis D\sigma_l }{\dis  \kappa_0'\ell_l(1-\sigma_l)}\delta_{l,k}+\frac{\dis  D\sigma_k }{\dis  N\kappa_0'(1-\sigma_l)\ell_k} .
\end{align}
Equation (\ref{matte}) then has the asymptotic expansion
\begin{align}
\label{matte2}
\pi_j(\x_0)& \sim (1-\sigma_j)\sum_{l,k=1}^N {\bf D}^{-1}_{jl} [\delta_{l,k}+\nu E_{lk}]\overline{\pi}_k(\x_0)\nonumber\\
&=\sum_{k=1}^N  [\delta_{j,k}+\nu E_{jk}]\overline{\pi}_k(\x_0)\nonumber \\
&=\overline{\pi}_j(\x_0)\bigg [1-\frac{\dis \nu D\sigma_l }{\dis  \kappa_0'\ell_j(1-\sigma_j)}\bigg ]+\frac{\dis \nu  D }{\dis  N\kappa_0'(1-\sigma_j)} \sum_{k=1}^N\frac{\sigma_k\overline{\pi}_k(\x_0)}{\ell_k}.
\end{align}
In the homogeneous case with $\sigma_j=\sigma_0$ and $\ell_j=\ell_0$ for all $j=1,\ldots,n$ we obtain the simple result
\begin{align}
\label{matte3}
\pi_j(\x_0)& \sim \overline{\pi}_j(\x_0)\bigg [1-\frac{\dis \nu D\sigma_0 }{\dis  \kappa_0'\ell_0(1-\sigma_0)}\bigg ]+\frac{\dis \nu  D \sigma_0}{\dis  N\kappa_0' \ell_0(1-\sigma_0)}.
\end{align}
That is, the leading order affect of desorption is to reduce (increase) all splitting probabilities larger (smaller) than $1/N$.

\section{Discussion} In this paper we considered a major extension of the narrow capture problem, in which finding a target $\calU$ by binding (adsorbing to) its surface $\partial \calU$ is not sufficient. In the case of animal foraging this could represent the need to access the resources within the interior of the target. Analogously, the binding of a signaling molecule to a surface receptor cannot initiate downstream processes until the receptor complex is internalized. Both examples require the important distinction between adsorption and absorption. The former represents the process whereby a diffusing particle succeeds in binding to the target surface, and is characterized by an effective binding rate or reactivity $\kappa_0$. Absorption, on the other hand, is the process of internalization which typically competes with unbinding or desorption from the surface.
We formulated the narrow capture problem for multiple partially accessible targets in terms of a set of renewal equations that relate the probability density and target fluxes in the presence of absorption to the corresponding quantities for irreversible adsorption. This allowed us to incorporate non-Markovian models of absorption and desorption, and to consider different desorption scenarios by modifying the rule for sewing together successive rounds of search-and-capture. In particular, rather than continuing the search process from the point of desorption, the particle could rapidly return to its initial position following desorption.
We solved the general renewal equations in two stages. First, we calculated the Laplace transformed target fluxes for irreversible adsorption by solving a Robbin BVP
 in the small-target limit using matched asymptotic analysis. Second, we used the inner solution of the BVP to solve the corresponding Laplace transformed renewal equations for non-Markovian desorption/absorption. This yielded explicit Neumann series expansions of the corresponding 
target fluxes that could be formally summed to yield expressions that are non-perturbative with respect to $\nu$. Finally, we derived expressions for the splitting probabilities and conditional mean FPTs for absorption by performing a small-$s$ expansion of the Neumann series.

There are a number of issues raised by this work that warrant further investigation. (i) The main focus of our study was to extend previous non-perturbative results for narrow capture problems to the case of partially accessible targets, see in particular equations (\ref{calJren}) and (\ref{calA2}). However, as a general practical tool for obtaining statistical quantities such as splitting probabilities and FPT moments, we will need to develop efficient procedures for extracting the small-$s$ behavior of the fluxes. In addition, it will be necessary to complement the asymptotic analysis with fast numerical schemes that can extend the results beyond the small-$\epsilon$ regime. (The summation over logarithmic singularities allows us to deal with the slow convergence of $\nu$ as $\epsilon \rightarrow 0$.)  (ii) It is also possible to incorporate non-Markovian models of adsorption. In Refs. \cite{Grebenkov23,Bressloff25} this is carried out for a single target using an encounter-based approach. The latter assumes that the probability of adsorption depends upon the amount of particle-surface contact time prior to binding as specified by the Brownian local time \cite{Grebenkov20,Bressloff22a}. (iii) Other obvious extensions include non-circular target shapes, which means replacing the rescaled radii $\ell_k$ by so-called shape capacitances, and the analogous 3D narrow capture problem. The details of the asymptotic analysis differs significantly in the latter case due to the $1/|\x-\x_0|$ singularity of the corresponding 3D Green's function $G(\x,s|\x_0)$ in the limit $\x\rightarrow \x_0$. (iv) Finally, one of the novel features of our analysis involves using the inner solution of a Robin BVP in the small target limit to eliminate surface integrals in a corresponding integral renewal equation. In general, dealing with such surface integrals requires finding an appropriate spectral decomposition of the integral kernels involving, for example, Dirichlet-to-Neumann operators defined on $L^2(\partial \calU)$ \cite{Grebenkov19a}. It would be interesting to explore other potential applications of our method.

\end{document}